\documentclass[twocolumn,showpacs,aps,superscriptaddress]{revtex4-1}
\usepackage{amsmath}
\usepackage{graphicx}
\usepackage{dcolumn}
\usepackage{float}
\newcommand{\be}{\begin{equation}}
\newcommand{\ee}{\end{equation}}
\newcommand{\bea}{\begin{eqnarray}}
\newcommand{\eea}{\end{eqnarray}}
\newcommand{\bs}{\begin{split}}
\newcommand{\bse}{\begin{subequations}}
\newcommand{\ese}{\end{subequations}}

\begin{document}
\title{Observation of long range magnetic ordering in pyrohafnate Nd$_2$Hf$_2$O$_7$: A neutron diffraction study}
\author{V. K. Anand}
\altaffiliation{vivekkranand@gmail.com}
\author{A. K. Bera}
\altaffiliation{Present Address: Solid State Physics Division, Bhabha Atomic Research Centre, Mumbai 400085, India}
\author{J. Xu}
\affiliation{\mbox{Helmholtz-Zentrum Berlin f\"{u}r Materialien und Energie GmbH, Hahn-Meitner Platz 1, D-14109 Berlin, Germany}}
\author{T. Herrmannsd\"orfer}
\affiliation{\mbox{Helmholtz-Zentrum Dresden-Rossendorf, Bautzner Landstrasse 400, D-01328 Dresden, Germany}}
\author{C. Ritter}
\affiliation{\mbox{Institut Laue-Langevin, Boite Postale 156, 38042 Grenoble Cedex, France}}
\author{B. Lake}
\altaffiliation{bella.lake@helmholtz-berlin.de}
\affiliation{\mbox{Helmholtz-Zentrum Berlin f\"{u}r Materialien und Energie GmbH, Hahn-Meitner Platz 1, D-14109 Berlin, Germany}}
\date{\today}

\begin{abstract}
We have investigated the physical properties of a pyrochlore hafnate Nd$_2$Hf$_2$O$_7$ using ac magnetic susceptibility $\chi_{\rm ac}(T)$, dc magnetic susceptibility  $\chi(T)$, isothermal magnetization $M(H)$ and heat capacity $C_{\rm p}(T)$ measurements, and determined the magnetic ground state by neutron powder diffraction study. An upturn is observed below 6~K in $C_{\rm p}(T)/T$, however both $C_{\rm p}(T)$ and $\chi(T)$ do not show any clear anomaly down to 2~K\@. The $\chi_{\rm ac}(T)$ shows a well pronounced anomaly indicating an antiferromagnetic transition at $T_{\rm N}= 0.55$~K\@. The long range antiferromagnetic ordering is confirmed by neutron diffraction. The refinement of neutron diffraction pattern reveals an all-in/all-out antiferromagnetic structure, where for successive tetrahedra, the four Nd$^{3+}$ magnetic moments point alternatively all-into or all-out-of the tetrahedron, with an ordering wavevector {\bf k} = (0, 0, 0) and an ordered state magnetic moment of $m = 0.62(1)\,\mu_{\rm B}$/Nd at 0.1~K\@. The ordered moment is strongly reduced reflecting strong quantum fluctuations in ordered state.
\end{abstract}


\maketitle

\section{\label{Intro} INTRODUCTION}

\begin{figure}
\includegraphics[width=2.6in, keepaspectratio]{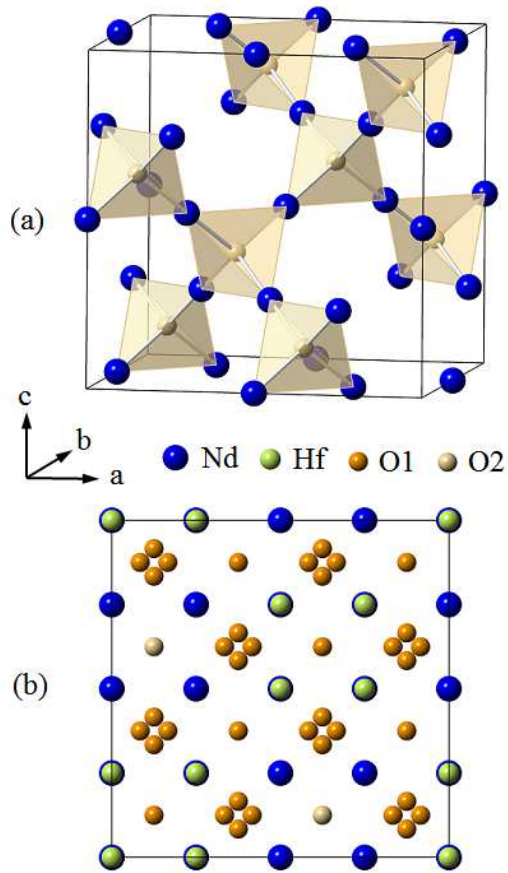}
\caption {(Color online) (a) An illustration of corner-sharing tetrahedra formed by Nd atoms for the face centered cubic (space group $Fd\bar{3}m$) pyrochlore structure of  Nd$_2$Hf$_2$O$_7$. (b) Projection of atomic arrangements onto the $ac$ plane.}
\label{fig:struct}
\end{figure}

The observations of spin-ice behavior and magnetic monopoles in frustrated 227 rare earth pyrochlores, such as in Dy$_2$Ti$_2$O$_7$ and Ho$_2$Ti$_2$O$_7$, have created tremendous research interests in these materials \cite{Gardner2010,Castelnovo2012,Gingras2014,Harris1997,Ramirez1999,Siddharthan1999,Hertog2000,Bramwell2001,Castelnovo2008,Morris2009,Jaubert2009,Bramwell2009,Fennell2009}.
In these pyrochlores the magnetic ions sit at the vertices of the corner-sharing tetrahedra (see Fig.~\ref{fig:struct}), and in this topology of the lattice, under the action of the crystal electric field (CEF), the ferromagnetic interaction between the moments becomes frustrated  \cite{Harris1997}. The spin-ice behavior of Dy$_2$Ti$_2$O$_7$ and Ho$_2$Ti$_2$O$_7$ is a nice illustration of such CEF dictated frustration of the ferromagnetic dipolar interaction.  It gives rise to an almost classical Ising spin system, forcing the magnetic moments at the corners of the tetrahedra to point along the local cubic $\langle 111 \rangle$ directions i.e., along the tetrahedron axes such that the moments can point only either towards the center or away from the center of each tetrahedra. Under these conditions the ferromagnetic exchange energy of an individual tetrahedron is minimized by ``two-in/two-out" spin configuration \cite{Harris1997} referred as the ``ice rule" by analogy with the disordered configurations of protons in water ice.  With this ``two-in/two-out" spin configuration, a Pauling residual entropy is found even in the $T = 0$ limit \cite{Ramirez1999}.

A delicate competition and balance between the magnetic exchange, CEF and dipolar interactions lead to a variety of very rich and unconventional low-temperature magnetic and thermodynamic properties in these frustrated 227 pyrochlore materials, for example observation of Dirac strings in spin ice Dy$_2$Ti$_2$O$_7$ \cite{Morris2009, Jaubert2009}, spin-liquid behavior in Tb$_2$Ti$_2$O$_7$ \cite{Gardner1999,Gardner2001,Molavian2007}, and a Higgs transition in quantum spin-ice Yb$_2$Ti$_2$O$_7$ \cite{Ross2011,Chang2012}. While the ferromagnetic dipolar interaction is frustrated by the Ising anisotropy, exchange interactions are antiferromagnetic and in the case they are stronger than the dipolar interactions, can result in a long range antiferromagnetic ordering as has been observed in several of 227 pyrochlore compounds \cite{Gardner2010,Hertog2000}.

The Heisenberg antiferromagnet Gd$_2$Ti$_2$O$_7$ exhibits long range magnetic ordering below $T_{\rm N} \approx 1.1$~K accompanied by another magnetic transition near 0.7~K and additional magnetic field induced transitions \cite{Raju1999,Ramirez2002,Stewart2004,Petrenko2004}. The $XY$-antiferromagnet Er$_2$Ti$_2$O$_7$ with the moments constrained to the local $\langle 111 \rangle$ planes orders below $T_{\rm N} \approx 1.2$~K \cite{Reotier2012} where the mechanism driving the ordering is suggested to be order-by-disorder \cite{Zhitomirsky2012,Ross2014}. The long range antiferromagnetic orderings of Ir$^{4+}$ ($5d^{5}$, $S=1/2$) moments in iridate pyrochlores $R_2$Ir$_2$O$_7$ ($R$ = Nd--Yb) have recently attracted attention for the associated metal-insulator transition \cite{Gardner2010,Krempa2014}. Eu$_2$Ir$_2$O$_7$ with Eu in Eu$^{3+}$ ($J=0$) exhibits an antiferromagnetic ordering accompanied with a metal-insulator transition at 120~K \cite{Zhao2011}. Magnetic structure determination using resonant x-ray diffraction revealed an all-in/all-out antiferromagnetic structure of  Ir$^{4+}$ in this compound with a propagation vector {\bf k} = (0, 0, 0) \cite{Sagayama2013}. In Nd$_2$Ir$_2$O$_7$ which exhibits a metal-insulator transition at 33~K, ordering of both Nd$^{3+}$ and Ir$^{4+}$ moments has been suggested by neutron diffraction (ND) and muon spin relaxation measurements with an all-in/all-out magnetic structure \cite{Tomiyasu2012,Guo2013}.

In our effort to search for novel 227 rare earth pyrochlores, we have investigated the physical properties of a pyrochlore hafnate Nd$_2$Hf$_2$O$_7$ having Nd$^{3+}$ ($4f^{3}$, $^4I_{9/2}$) as magnetic ion with $S=3/2$, $L=6$ and $J=9/2$. This compound has recently been studied for its promising dielectric properties (high dielectric constant) \cite{Wei2009, Chun2015}, however its magnetic properties have not been investigated. Ubic {\it et al}.\ \cite{Ubic2008} suggested presence of small disorder in oxygen sublattice of Nd$_2$Hf$_2$O$_7$, however, latter investigations by Karthik {\it et al}.\ \cite{Karthik2012} revealed a well-ordered pyrochlore structure. Our x-ray and neutron powder diffraction data confirm the ordered cubic $Fd\bar{3}m$  pyrochlore structure.

Here we report the results of ac magnetic susceptibility $\chi_{\rm ac}$, dc magnetic susceptibility  $\chi$, isothermal magnetization $M$ and heat capacity $C_{\rm p}$ measurements on Nd$_2$Hf$_2$O$_7$ as a function of temperature $T$ and magnetic field $H$. The $\chi_{\rm ac}(T)$ data which is measured down to 0.2~K show evidence for an antiferromagnetic transition at $T_{\rm N}= 0.55$~K\@. The antiferromagnetic long range ordering is further confirmed by neutron powder diffraction which reveals an all-in/all-out magnetic structure with an ordered moment of $0.62(1)\,\mu_{\rm B}$/Nd at 0.1~K and propagation wavevector {\bf k} = (0, 0, 0). Very recently Nd$_2$Zr$_2$O$_7$ has also been identified to order antiferromagnetically below 0.3~K with an all-in/all-out magnetic structure \cite{Lhotel2015,Xu2015}. Investigations on the single crystal Nd$_2$Zr$_2$O$_7$ have revealed local $\langle 111 \rangle$ Ising anisotropy \cite{Hatnean2015} for the ground state of Nd$^{3+}$. Our $M(H)$ and $\chi(T)$ data consistently comply with the Ising nature of ground state of Nd$^{3+}$ in Nd$_2$Hf$_2$O$_7$ with an effective spin $S = 1/2$ and effective $g$-factor $g_{zz} = 5.01(3)$.

An interesting aspect of the rare earth pyrochlores with Kramers doublet (like Nd$^{3+}$) having well separated ground state and first excited state is that the ground state properties can be described by an effective pseudo-spin $S = 1/2$. Of particular interests are the Kramers doublet systems with total angular momenta $J = 9/2$ (Nd$^{3+}$) and 15/2 (Dy$^{3+}$ and Er$^{3+}$) for which Huang {\it et al}.\ \cite{Huang2014} showed that under specific conditions they may behave like `dipolar-octupolar' doublets. For a dipolar-octupolar doublet the $x$ and $z$ components of pseudo-spin operator transform like the $z$ component of a magnetic dipole whereas the $y$ component transforms like a component of a magnetic octupolar tensor \cite{Huang2014}. The fact that the pseudo-spin ground state of Nd$^{3+}$ with $J=9/2$ can behave like dipolar-octupolar doublet makes Nd$_2$Hf$_2$O$_7$ an interesting system for further investigations.

\section{\label{ExpDetails} Experimental Details}

A polycrystalline sample of Nd$_2$Hf$_2$O$_7$ was synthesized by solid state reaction method using Nd$_2$O$_3$ (99.99\%) and HfO$_2$ (99.95\%) from Alfa Aesar. A stoichiomteric mixture of finely ground Nd$_2$O$_3$ and HfO$_2$ was first fired at 1300~$^{\circ}$C for 50~h, with two more succesive grindings and firings at 1400~$^{\circ}$C and 1500~$^{\circ}$C, each for 50~h\@. The finely ground mixture was then pressed into a pellet and fired at 1550~$^{\circ}$C for 80~h\@. We also synthesized the nonmagnetic reference compound La$_2$Hf$_2$O$_7$ using La$_2$O$_3$ (99.999\%, Alfa Aesar) and HfO$_2$ with similar grinding and firing sequence however the last firing after pelletizing was done at 1500~$^{\circ}$C\@. An agate mortar and pestle was used for grinding with a grinding of approximately half an hour to achieve homogeneity, and a zirconium oxide crucible was used for firing. The grinding and heat treatment of the samples were done in air.

The crystal structure and the quality of the samples were checked by room temperature powder x-ray diffraction (XRD, Brucker). A commercial superconducting quantum interference device (SQUID) magnetometer (MPMS, Quantum Design Inc.) was used for dc magnetic measurements at Mag Lab, Helmholtz-Zentrum Berlin (HZB). The ac susceptibility was collected at temperatures down to 200 mK using an adiabatic demagnetization cooler where the sample is located in the center of mutual inductance. The ac susceptibility was measured using a LR700 mutual inductance bridge. As a magnetocaloric active substance for the sub-K temperature range, an Fe$^{3+}$ salt was used. The demagnetization stage was precooled in a physical property measurement system (PPMS, Quantum Design Inc.) at Helmholtz-Zentrum Dresden-Rossendorf.  Heat capacity was also measured using PPMS by means of the adiabatic-relaxation technique down to 2~K at Mag Lab, HZB.

The neutron diffraction measurements were carried out using the D20 powder neutron diffractometer at the Institute Laue Langevin, Grenoble, France. A thin-walled copper can (diameter 10~mm) was used to mount the powdered sample. Low temperatures down to 0.1~K were achieved by cooling the sample in a dilution fridge. High intensity ND data were collected at few selected temperatures between 0.1~K and 0.6~K\@.  Incident neutron beam of wavelength $\lambda = 2.41$~\AA\ was used for these measurements and counted for 4 hours at each of the temperatures. The XRD and ND data were refined using the package FullProf suite \cite{Rodriguez1993}.

\section{\label{Crystallography} Crystallography}

\begin{figure}
\includegraphics[width=3.2in, keepaspectratio]{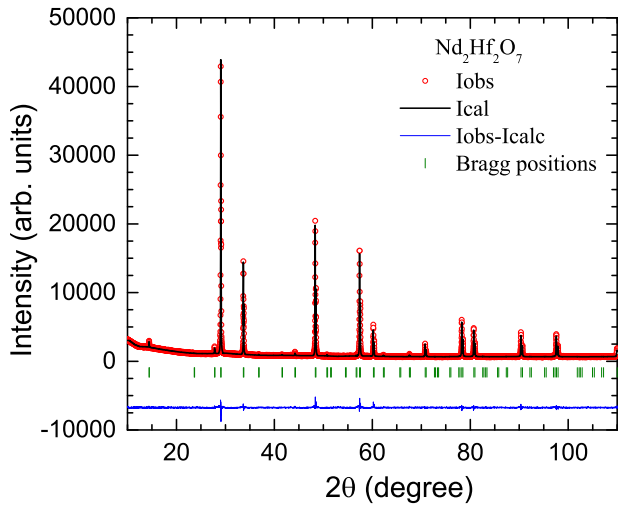}
\caption {(Color online) Powder x-ray diffraction pattern of Nd$_2$Hf$_2$O$_7$ recorded at room temperature. The solid line through the experimental points is the Rietveld refinement profile calculated for the  ${\rm Eu_2Zr_2O_7}$-type face centered cubic (space group $Fd\bar{3}m$) pyrochlore structure. The short vertical bars mark the Bragg-peak positions. The lowermost curve represents the difference between the experimental and calculated intensities.}
\label{fig:XRD}
\end{figure}

Figure~\ref{fig:XRD} shows the Rietveld refinement of powder XRD pattern of Nd$_2$Hf$_2$O$_7$ collected at room temperature. The refinement confirms the ${\rm Eu_2Zr_2O_7}$-type face-centered cubic (space group $Fd\bar{3}m$) pyrochlore structure of Nd$_2$Hf$_2$O$_7$. All the observed peaks are indexed, thus revealing the single phase nature of the sample. The crystallographic parameters obtained from the structural refinement of the room temperature XRD are listed in Table~\ref{tab:XRD} along with those obtained from the refinement of the neutron diffraction data recorded at 0.6~K\@. The parameters $a$ and $x_{\rm O1}$ agree well with the literature values \cite{Ubic2008, Karthik2012}. Because of the limitations of laboratory based x-ray measurements it is not possible to obtain any reliable information about the oxygen vacancy or Nd-Hf site mixing. Our neutron data which were collected at low Q are also of not much help in resolving these issues because of the small number of nuclear Bragg peaks detected and almost equal scattering lengths for Nd ($0.7690 \times 10^{-12}$~cm) and Hf ($0.7770\times 10^{-12}$~cm).

\begin{table}
\caption{\label{tab:XRD} Refined crystallographic parameters and agreement factors obtained from the structural Rietveld refinement of room temperature (RT) powder XRD data and 0.6~K neutron powder diffraction data for Nd$_2$Hf$_2$O$_7$. The Wyckoff positions of Nd, Hf, O1 and O2 atoms in space group $Fd\bar{3}m$ are 16d (1/2,1/2,1/2), 16c (0,0,0), 48f ($x_{\rm O1}$,1/8,1/8) and 8b (3/8,3/8,3/8), respectively. The atomic coordinate $x_{\rm O1}$ is listed below.}
\begin{ruledtabular}
\begin{tabular}{lcc}
 & XRD (RT) & ND (0.6~K)\\
 \hline
 \underline{Lattice parameters}\\
{\hspace{0.8cm} $a$ ({\AA})}            			&  10.6389(1) & 10.573(1) \\	
{\hspace{0.8cm} $V_{\rm cell}$  ({\AA}$^{3}$)} 	&  1204.19(1)  & 1182.0(2) \\
\underline{Atomic coordinate}\\
\hspace{0.8cm} $x_{\rm O1}$ & 0.3317(9) & 0.3340(5)\\
\underline{Refinement quality} \\
\hspace{0.8cm} $\chi^2$   & 2.86 & 1.98\\	
\hspace{0.8cm} $R_{\rm p}$ (\%)  & 3.69 &  5.97\\
\hspace{0.8cm} $R_{\rm wp}$ (\%) & 5.27 & 10.70 \\
\hspace{0.8cm} $R_{\rm Bragg}$ (\%) & 5.19 & 6.48 \\
\end{tabular}
\end{ruledtabular}
\end{table}

The pyrochlore structure of Nd$_2$Hf$_2$O$_7$ is shown in Fig.~\ref{fig:struct}. In this structure the Nd atoms form corner-shared tetrahedra, as shown in Fig.~\ref{fig:struct}(a), the center of each tetrahedron is occupied by an O atom. The atomic arrangements viewed along the crystallographic $b$ direction (projected in $ac$ plane) are shown in Fig.~\ref{fig:struct}(b). The Nd$^{3+}$ occupy 16d (1/2,1/2,1/2) sites and the Hf$^{4+}$ occupy 16c (0,0,0) positions, whereas the O$^{2-}$  occupy two sites: O1 in 48f ($x_{\rm O1}$,1/8,1/8) and O2 in 8b (3/8,3/8,3/8) positions, and the formula unit can be viewed as ${\rm Nd_2Hf_2O(1)_6O(2)}$. The atoms sitting at both the 16d and 16c sites form (separately) three-dimensional networks of corner-shared tetrahedra leading to two distinct pyrochlore sublattices [only the tetrahedra formed by 16d site atoms (Nd here) are shown in Fig.~\ref{fig:struct}(a)]. The Nd atoms are eight-fold coordinated (by 6 O1 and 2 O2) and the Hf atoms are six-fold cordinated (by 6 O1) \cite{Subramanian1983}.

The La$_2$Hf$_2$O$_7$ also forms in the same face-centered cubic pyrochlore structure with parameters $a = 10.7731(1)$~\AA\ and $x_{\rm O1} = 0.3308(9)$ in agreement with the reported values \cite{Karthik2012}. The single phase nature of sample was inferred from the refinement of XRD data (not shown).

\section{\label{Sec:ChiMH} Magnetization and Magnetic Susceptibility}

\begin{figure}
\includegraphics[width=3in, keepaspectratio]{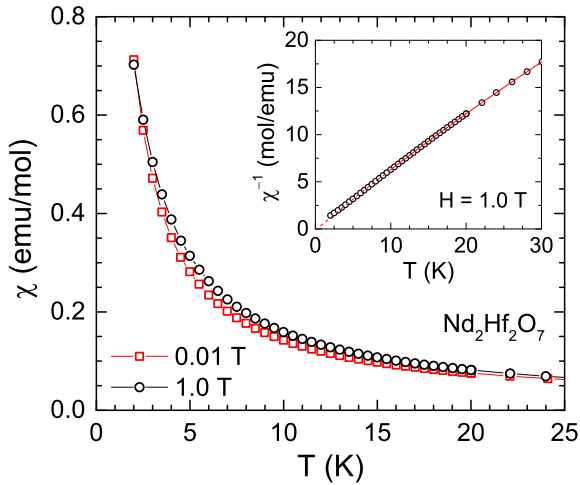}
\caption{(Color online) Zero-field-cooled magnetic susceptibility $\chi$ of Nd$_2$Hf$_2$O$_7$ as a function of temperature $T$ for $2~{\rm K} \leq T \leq 25$~K measured in magnetic fields $H= 0.01$~T and 1.0~T\@. Inset: Inverse magnetic susceptibility $\chi^{-1}(T)$ for $2~{\rm K} \le T \leq 30$~K in $H = 1.0$~T\@. The solid red line is the fit of the $\chi^{-1}(T)$ data by the Curie-Weiss law in $10~{\rm K} \leq T \leq 30$~K and the dashed line is an extrapolation. All data pertain to per mole of Nd$_2$Hf$_2$O$_7$.}
\label{fig:MT}
\end{figure}

Zero-field-cooled dc $\chi(T)$ data of Nd$_2$Hf$_2$O$_7$ measured in $H = 0.01$~T and 1.0~T are shown in Fig.~\ref{fig:MT}. The $\chi(T)$ data do not show any anomaly and remain paramagnetic at $T \geq 2$~K\@. The high temperature $\chi(T)$ data follow Curie-Weiss behavior, $\chi(T) = C/(T-\theta_{\rm p})$. The linear fit of $\chi^{-1}(T)$ for the range 100~K~$\leq T\leq$~300~K gives the Curie constant $C = 1.32(1)$~emu\,K/mol\,Nd and Weiss temperature $\theta_{\rm p} = -29.9(7)$~K\@. The $C$ value gives an effective moment $\mu_{\rm eff} = 3.25\, \mu_{\rm B}$/Nd according to the relation $C= N_{\rm A} \mu_{\rm eff}^2/3 k_{\rm B}$ where $N_{\rm A}$ is the Avogadro number and $k_{\rm B}$ is the Boltzmann constant. The obtained $\mu_{\rm eff}$ is little smaller than the theoretically expected value of effective moment for $^4$I$_{9/2}$ ground state of Nd$^{3+}$ ions ($\mu_{\rm eff} = g_J\sqrt{J(J+1)} = 3.62 \, \mu_{\rm B}$ for $g_J = 8/11$ and $J = 9/2$).

As estimated in the next section, the first excited crystal field level is situated at about 230 K, therefore because of the thermal population from CEF at $T \geq 100$~K, the above analysis of $\chi(T)$ data does not give the correct estimate of $\theta_{\rm p}$ or $\mu_{\rm eff}$ for the Ising ground state. Therefore we fit the $\chi(T)$ data at $T$ below 30~K using the modified Curie-Weiss behavior  $\chi(T) = \chi_0 + C/(T-\theta_{\rm p})$ where temperature independent term $\chi_0$ is added to account for the Van Vleck contribution. The fit of $\chi(T)$ by this modified Curie-Weiss law is shown by solid line in the inset of Fig.~\ref{fig:MT} plotted as the inverse of susceptibility $\chi^{-1}(T)$.  In order to avoid the effect of short-range magnetic interactions at $T < 10$~K as evident from the heat capacity data discussed in next section, we fit the $\chi^{-1}(T)$ data in 10~K~$\leq T\leq$~30~K which yields $\chi_0 = 3.04(4) \times 10^{-3}$~emu/mol\,Nd, $C = 0.752(2)$~emu\,K/mol\,Nd and $\theta_{\rm p} = +0.17(2)$~K\@. When the $\chi(T)$ data are corrected for demagnetization effects by roughly approximating the sample to be spherical in shape we obtain $\chi_0 = 3.04(4) \times 10^{-3}$ emu/mol\,Nd, $C = 0.753(2)$~emu\,K/mol\,Nd and $\theta_{\rm p} = +0.24(2)$~K\@. Thus we see that $\theta_{\rm p}$ is positive which would imply a weak ferromagnetic coupling among Nd spins. A positive $\theta_{\rm p}$ was also found in the case of Nd$_2$Zr$_2$O$_7$ \cite{Xu2015,Hatnean2015}. The $C$ value gives $\mu_{\rm eff} \approx 2.45\, \mu_{\rm B}$/Nd for the Ising ground state of Nd$_2$Hf$_2$O$_7$.

\begin{figure}
\includegraphics[width=3in, keepaspectratio]{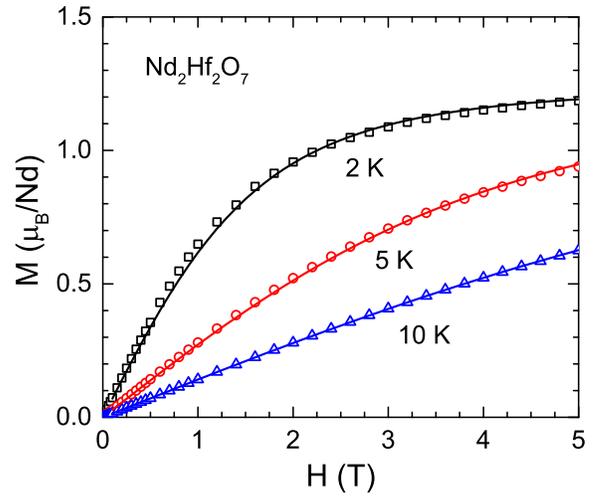}
\caption{(Color online) Isothermal magnetization $M$ (per Nd ion) of Nd$_2$Hf$_2$O$_7$ as a function of applied magnetic field $H$ for $0 \leq H \leq 5$~T measured at the indicated temperatures. The solid curves are the fits of $M(H)$ data by Eq.~(\ref{MH-Ising}) with an effective longitudinal $g$-factor $g_{zz}= 5.01(3)$.}
\label{fig:MH}
\end{figure}

The isothermal $M(H)$ data of Nd$_2$Hf$_2$O$_7$ at $T= 2$~K, 5~K and 10~K are shown in Fig.~\ref{fig:MH}. At 2~K initially $M$ increases rapidly and is linear in $H$ for $H\leq 0.5$~T above which $M(H)$ shows nonlinear behavior and tends towards saturation with a magnetization value of $M \approx 1.2\, \mu_{\rm B}$/Nd at 5~T which is much lower than the theoretical saturation magnetization $M_{\rm sat} = g_J J\,\mu_{\rm B} = 3.27\,\mu_{\rm B}$/Nd for free Nd$^{3+}$ ions ($g_J = 8/11$ and $J = 9/2$). The observed $M$ is only about 37\% of the free ion theoretical saturation value and reflects substantial single-ion anisotropy as would one expect for a local $\langle 111 \rangle$ Ising anisotropic system. With increasing $T$ the linear regime of $M(H)$ extends over a large field range, although at a more gradual rate.

As the first excited state is well separated ($\sim230$~K) from the ground state, the low temperature magnetic properties of Kramers ground doublet of Nd$^{3+}$  can be described by an effective spin $S = 1/2$. For a Kramers doublet of dipolar-octupolar type (Nd$^{3+}$) the transverse $g$-factor is found to be zero, i.e.\ $g_{\bot} = 0$. For an effective spin-half doublet ground state system with local $\langle 111 \rangle$ Ising anisotropy assuming $g_{\bot} = 0$ and $g_{||}=g_{zz}$ the powder- and thermally-averaged magnetization is given by  \cite{Bramwell2000}
\begin{equation}
\langle M \rangle = \frac{(k_{\rm B} T)^2}{g_{zz}\mu_{\rm B} H^2 S} \int_0^{g_{zz}\mu_{\rm B} H S/k_{\rm B} T} x \tanh(x) \,dx
\label{MH-Ising}
\end{equation}
where $x = g_{zz}\mu_{\rm B} H S/k_{\rm B} T$. For a pure $m_J = \pm 9/2$ doublet $g_{zz} = 2 g_J J = 6.54$. However, due to the mixing of the $m_J$ states by the crystal field the effective $g$-factor is different from 6.54 and can be determined by fitting the $M(H)$ data which is the only adjustible parameter in Eq.~(\ref{MH-Ising}). The fits of $M(H)$ data at $T= 2$~K, 5~K and 10~K (solid curves in Fig.~\ref{fig:MH}) yield $g_{zz}= 5.01(3)$. It is seen from Fig.~\ref{fig:MH} that the $M(H)$ data are reasonably well described by Eq.~(\ref{MH-Ising}).  The $g_{zz} = 5.01(3)$ obtained this way is lower than that of a pure $m_J = \pm 9/2$ doublet. This reduction of $g_{zz}$ possibly suggests an admixture of other $m_J$ terms in the ground state \cite{Bramwell2000}. For Nd$_2$Zr$_2$O$_7$ $g_{zz}$ is found to be 4.793 \cite{Hatnean2015} and 5.30(6) \cite{Xu2015}. The effective $g_{zz} = 5.01(3)$ with an effective $S = 1/2$ suggests an Ising moment of $g_{zz} S  \, \mu_{\rm B} = 2.50 \, \mu_{\rm B}$/Nd. For a powder sample the effective moment is related to $g$-factor as $\mu_{\rm eff} = (\sqrt 3/2)\overline{g}\,\mu_{\rm B} $, where $\overline{g}^2 = (g_{||}^2+2 g_{\bot}^2)/3$, which for $g_{\bot} = 0$ and  $g_{||}= 5.01(3)$ gives $\mu_{\rm eff} = 2.50 \, \mu_{\rm B}$/Nd in agreement with the above inferred value of $ 2.45\,  \mu_{\rm B}$/Nd from the fit of $\chi(T)$ data. Thus the $M(H)$ and $\chi(T)$ data consistently follow the Ising behavior.

\begin{figure}
\includegraphics[width=3.1in, keepaspectratio]{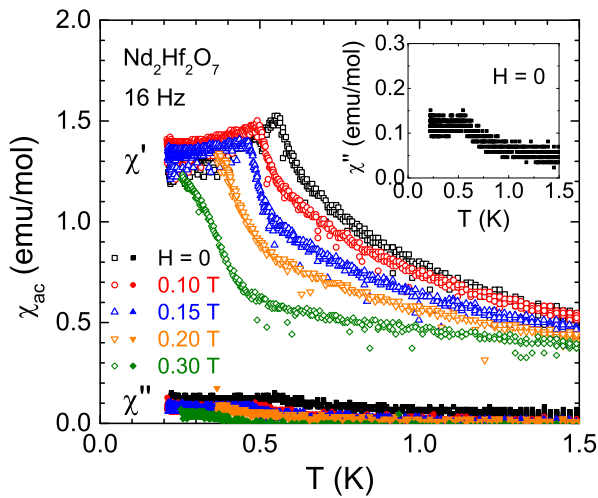}
\caption {(Color online) The temperature $T$ dependence of real $\chi'$ and imaginary $\chi''$ parts of ac magnetic susceptibility $\chi_{\rm ac}$ of Nd$_2$Hf$_2$O$_7$ measured in different dc magnetic fields at 16~Hz. Inset: An expanded view of $\chi''(T)$ in $H=0$. All data pertain to per mole of Nd$_2$Hf$_2$O$_7$. Because of the uncertainty in calculating the sample filling factor in the ac coil-set,  the conversion to units of `emu/mol' for $y$-scale has accuracy of about 10\%.}
\label{fig:Chiac}
\end{figure}

The real $\chi'$ and imaginary $\chi''$ parts of ac magnetic susceptibility $\chi_{\rm ac}$ of Nd$_2$Hf$_2$O$_7$ measured in $H\leq 0.30$~T and 16~Hz are shown in Fig.~\ref{fig:Chiac} for low temperatures $T\leq 1.5$~K\@. In $H = 0$ the $\chi'(T)$ data show a pronounced peak at 0.55~K indicating a magnetic phase transition. Further, with increasing field the peak position shifts towards lower temperatures. This behavior is a characteristic of an antiferromagnetic phase transition. A weak anomaly with a similar $H$ dependence is also observed in imaginary part of ac susceptibility $\chi''(T)$. The $\chi''(T)$ is much smaller in magnitude than the $\chi'(T)$. Due to weak signal at 16~Hz the signal-to-noise ratio for $\chi''(T)$ is very poor and data appear quite noisy. As can be seen from the inset of Fig.~\ref{fig:Chiac}, despite the noisy data an anomaly near 0.55 K is also visible in the $\chi''(T)$ data in $H=0$. Thus the $\chi_{\rm ac}(T)$ data indicate a long range antiferrromagnetic ordering of Nd$^{3+}$ at the N\'eel temperature $T_{\rm N} = 0.55$~K as is confirmed by the neutron diffraction study discussed below in Sec.~\ref{Sec:ND}. Further it is seen that the application of field also causes a decrease in $\chi'$ at $T > T_{\rm N}$ likely due to the effect of field on short range magnetic correlations above $T_{\rm N} $.

\section{\label{Sec:HC}Heat Capacity}

\begin{figure}
\includegraphics[width=3in, keepaspectratio]{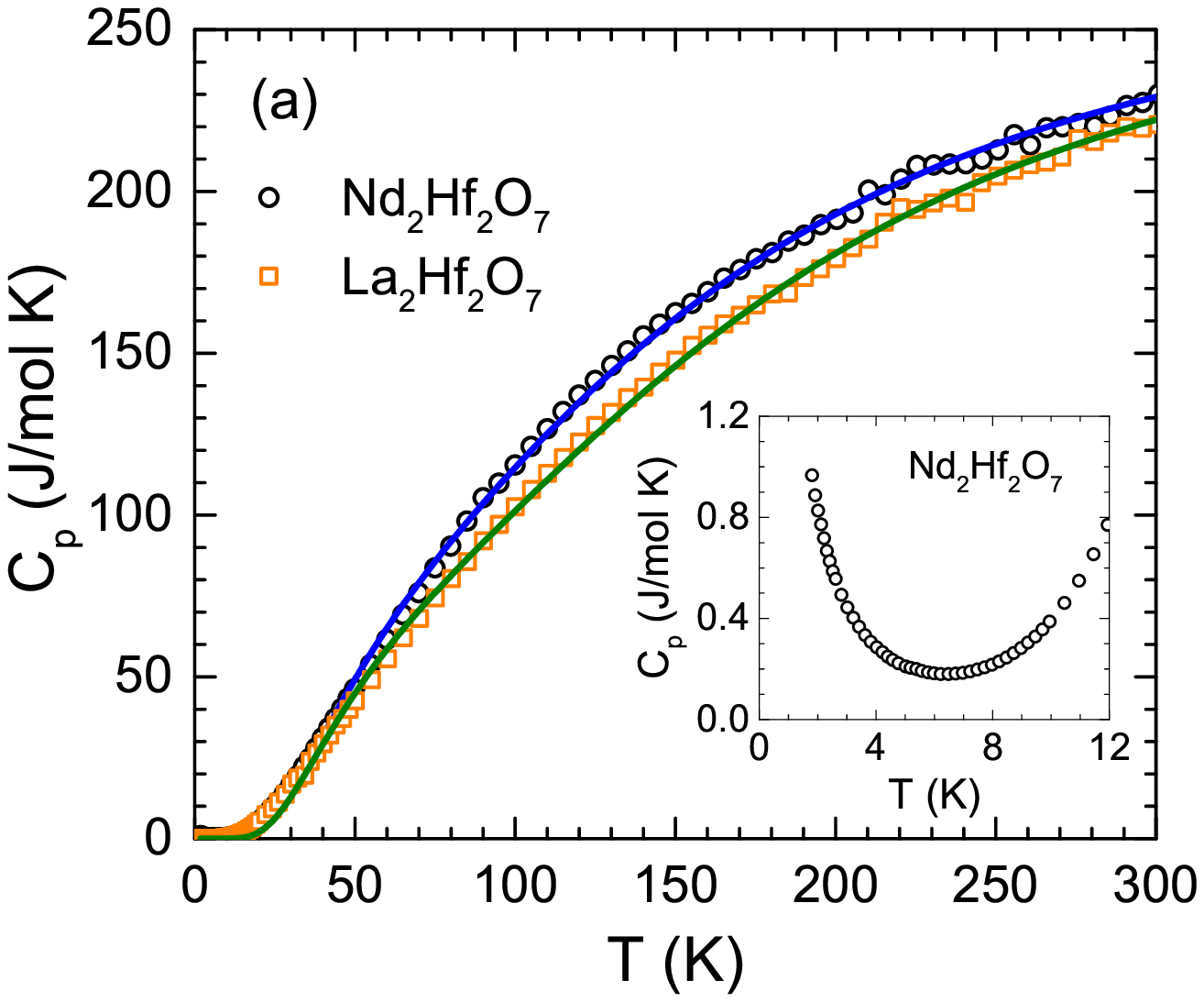}
\includegraphics[width=3in, keepaspectratio]{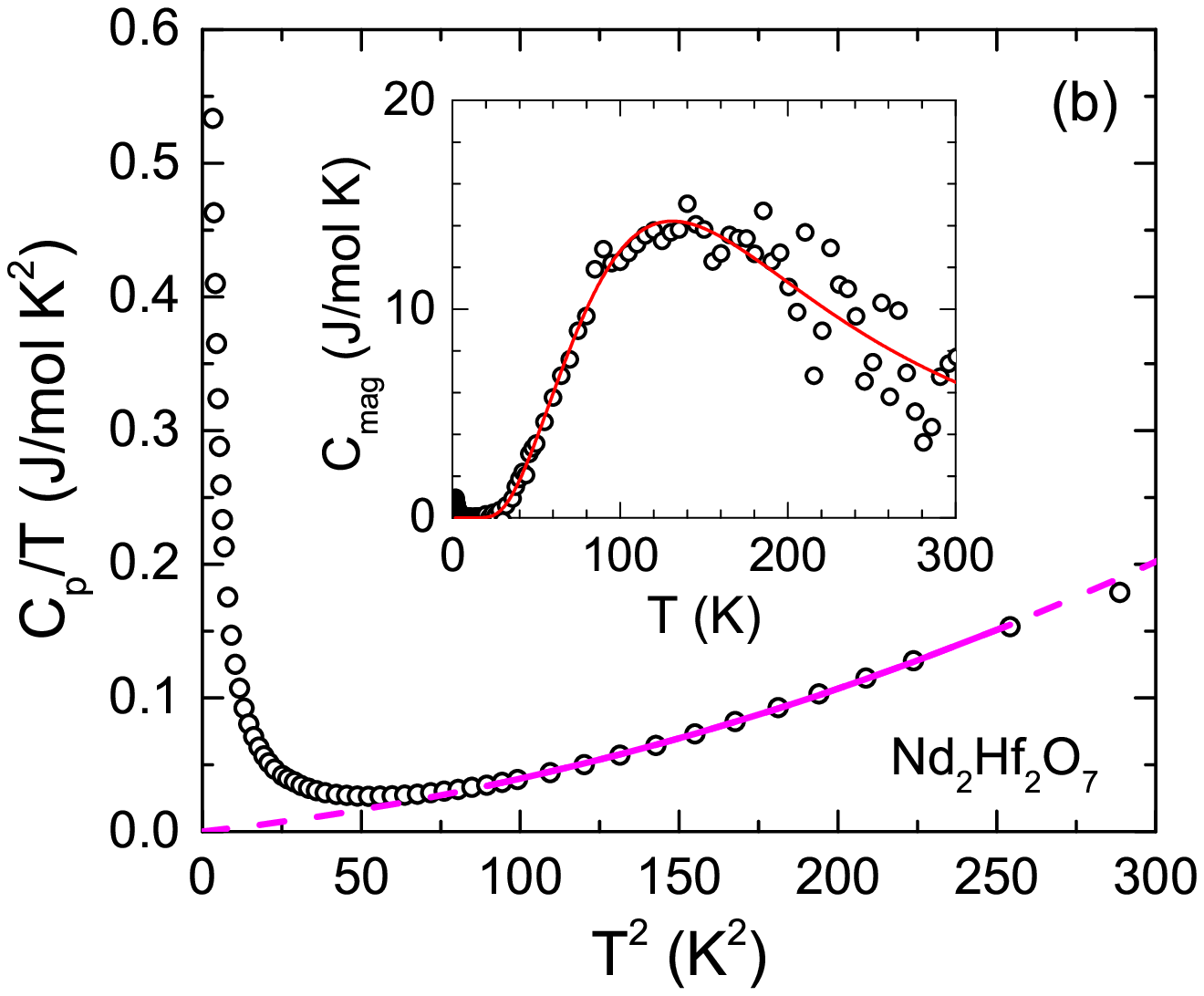}
\caption{\label{fig:HC} (Color online) (a) Heat capacity $C_{\rm p}$ of Nd$_2$Hf$_2$O$_7$ and nonmagnetic reference La$_2$Hf$_2$O$_7$ as a function of temperature $T$ for 1.8~K~$\leq T \leq$~300~K measured in zero field. The solid curves are the fits by Debye+Einstein models of lattice heat capacity (plus crystal field contribution for the case of Nd$_2$Hf$_2$O$_7$). Inset: Expanded view of \mbox{low-$T$} $C_{\rm p}(T)$ data over $1.8~{\rm K} \leq T \leq 12$~K\@ for Nd$_2$Hf$_2$O$_7$. (b) $C_{\rm p}/T$ versus $T^2$ plot for Nd$_2$Hf$_2$O$_7$ for $T \leq 17$~K\@. The solid line is the fit to $C_{\rm p}/T = \beta T^2 +\delta T^4 $ in $10~{\rm K} \leq T \leq 16$~K and the dashed lines are extrapolations. Inset: Magnetic contribution to heat capacity $C_{\rm mag}(T)$ for Nd$_2$Hf$_2$O$_7$. The solid curve represents the crystal field contribution to heat capacity as discussed in text.}
\end{figure}

The $C_{\rm p}(T)$ data of Nd$_2$Hf$_2$O$_7$ and nonmagnetic reference compound La$_2$Hf$_2$O$_7$ are shown in Fig.~\ref{fig:HC}(a) for 1.8~K~$\leq T \leq$~300~K\@. Consistent with the $\chi(T)$ data, the $C_{\rm p}(T)$ data of  Nd$_2$Hf$_2$O$_7$ do not show any anomaly at $T\geq1.8$~K. However, the low-$T$ $C_{\rm p}(T)$ at $T \leq 6$~K reveals an upturn as shown in the inset of Fig.~\ref{fig:HC}(a). This reflects the onset of short range magnetic correlation well above the antiferromagnetic transtion at $T_{\rm N} = 0.55$~K\@. The low-$T$ $C_{\rm p}(T)$ data (above 6~K) are well described by $C_{\rm p}(T) = \gamma T + \beta T^{3} + \delta T^{5}$, with the coefficient of electronic heat capacity $\gamma = 0$ which reflects an insulating ground state in Nd$_2$Hf$_2$O$_7$. A fit to the $C_{\rm p}/T$ versus $T^2$ plot by $C_{\rm p}/T = \beta T^2 +  \delta T^{4}$  over  $9.5~{\rm K} \leq T \leq 16$~K as shown by the solid magneta line in Fig.~\ref{fig:HC}(b) gives $\beta= 2.58(5) \times 10^{-4}$~J/mole\,K$^{4}$ and $\delta = 1.38(3) \times 10^{-6}$~J/mole\,K$^{6}$. The Debye temperature $\Theta_{\rm D} = 436(4)$~K is estimated from $\beta$ using the relation $\Theta_{\rm D} = (12 \pi^{4} n R /{5 \beta} )^{1/3}$ where $n=11$ is the number of atoms per formula unit and $R$ is the molar gas constant.

The $C_{\rm p} \approx 230$ J/mol\,K at 300~K [Fig.~\ref{fig:HC}(a)] is much lower than the expected high-$T$ limit Dulong-Petit value $C_{\rm V} = 3nR = 33R \approx 274.4$~J/mol\,K which is consistent with the high $\Theta_{\rm D}$ value. The $\Theta_{\rm D}$ in 227 pyrochlore is found to be highly temperature dependent, for Dy$_2$Ti$_2$O$_7$ the low-$T$ $C_{\rm p}(T)$ yields a $\Theta_{\rm D}$ of 295~K whereas the high-$T$ $C_{\rm p}(T)$ gives much higher $\Theta_{\rm D} = 722(8)$~K \cite{Anand2015a}. A better estimate of $\Theta_{\rm D}$ can be obtained from fitting the $C_{\rm p}(T)$ data by a combination of the Debye and Einstein models of lattice heat capacity. A fit of $C_{\rm p}(T)$ data of La$_2$Hf$_2$O$_7$ by Debye+Einstein models in 1.8~K~$\leq T \leq$~300~K gives $\Theta_{\rm D} = 792(5)$~K and Einstein temperature $\Theta_{\rm E} = 163(2)$~K. The fit is shown by the solid olive curve in Fig.~\ref{fig:HC}(a) which is obtained with 66\% weight to Debye term and 34\% to Einstein term. Further details about fitting heat capacity data by Debye+Einstein models can be found in Ref.~\cite{Anand2015a}.

On the other hand for Nd$_2$Hf$_2$O$_7$  we have an additional magnetic contribution due to crystal electric field. The magnetic contribution to heat capacity $C_{\rm mag}(T)$ for Nd$_2$Hf$_2$O$_7$ is shown in the inset of Fig.~\ref{fig:HC}(b)  which was obtained by subtracting the lattice heat capacity (equivalent to $C_{\rm p}(T)$ of La$_2$Hf$_2$O$_7$) from the heat capacity of Nd$_2$Hf$_2$O$_7$. A correction for the small formula mass difference of the two compounds was employed \cite{Anand2015b}. As can be seen from the inset of Fig.~\ref{fig:HC}(b), $C_{\rm mag}(T)$ is noisy at high-$T$  (due to the limitations in the sensitivity of our experimental setup), nevertheless the basic feature of the data is quite evident.  A broad Schottky type anomaly (due to crystal electric field) centered around 120~K is seen in $C_{\rm mag}(T)$. We analyzed $C_{\rm mag}(T)$ to extract the CEF levels and found that $C_{\rm mag}(T)$ is well represented by a doublet ground state lying below a doublet excited state at 229(6)~K and a quasi-quartet (two closely situated doublets) at 460(9)~K\@. The crystal field contribution to heat capacity calculated according to this CEF level scheme is shown by the solid red curve in the inset of  Fig.~\ref{fig:HC}(b). A nice agreement is seen between $C_{\rm mag}(T)$ and the CEF model fit. However, because of large noise in high-$T$ data the precise determination of the splitting energy for higher excited states is not possible. For Nd$_2$Zr$_2$O$_7$ a splitting energy of 250--270~K between the ground state doublet and first excited doublet has been found \cite{Hatnean2015,Xu2015,Lhotel2015}. Similar to the present compound, two closely situated doublets at 388~K and 400~K have also been found from inelastic neutron scattering study on  Nd$_2$Zr$_2$O$_7$ \cite{Lhotel2015}.

The fit of $C_{\rm p}(T)$ data of Nd$_2$Hf$_2$O$_7$ by CEF+Debye+Einstein models in 1.8~K~$\leq T \leq$~300~K shown by the solid blue curve in Fig.~\ref{fig:HC}(a) yields $\Theta_{\rm D} = 785(6)$~K and Einstein temperature $\Theta_{\rm E} = 162(2)$~K with 66\% weight to Debye term and 34\% to Einstein term. The value of $\Theta_{\rm D} = 785(6)$~K obtained this way is much higher than the $\Theta_{\rm D} = 436(4)$~K estimated above from $\beta$.

\section{\label{Sec:ND}Neutron Diffraction}

\begin{figure}
\includegraphics[width=3.1in, keepaspectratio]{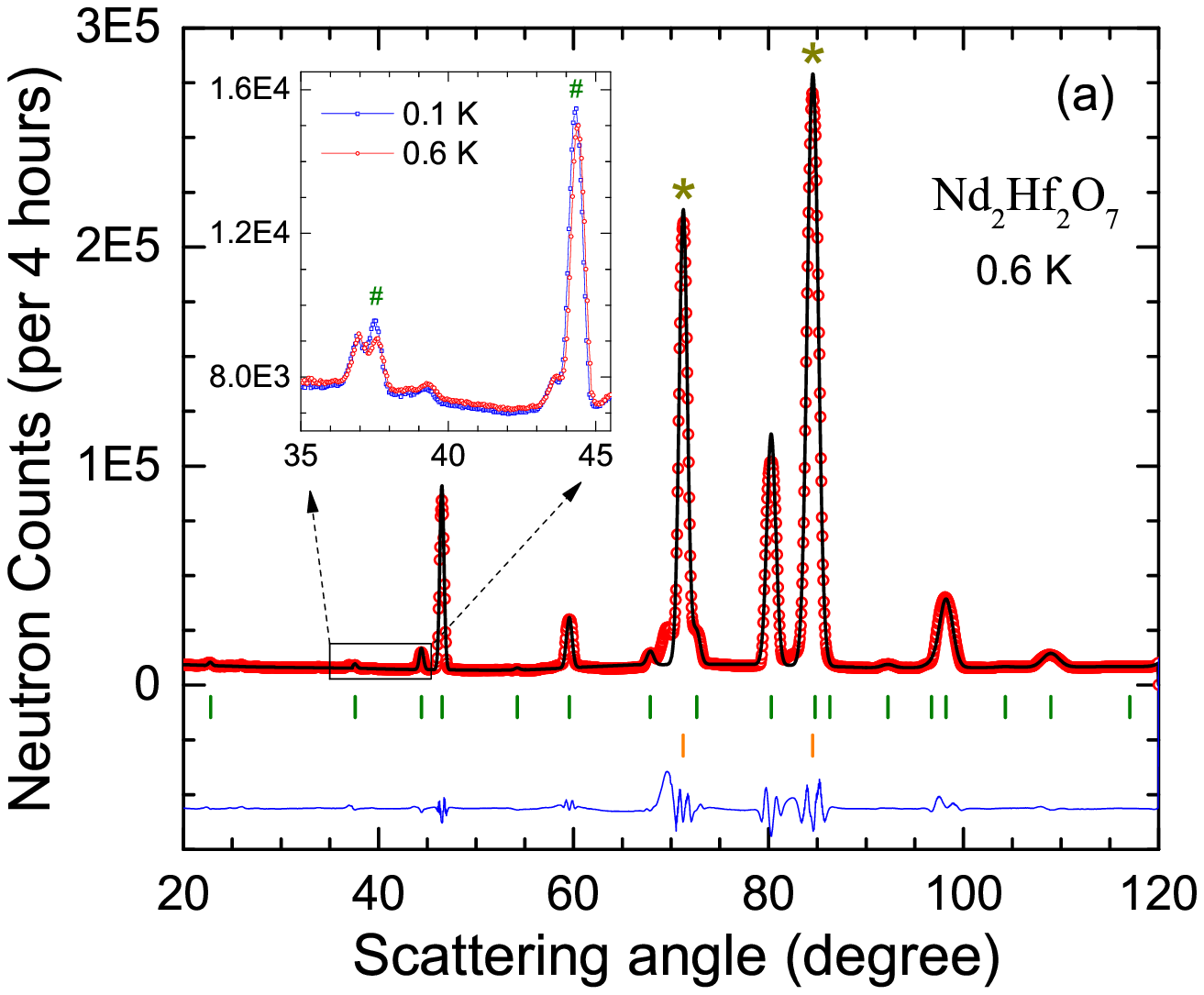}
\includegraphics[width=3.1in, keepaspectratio]{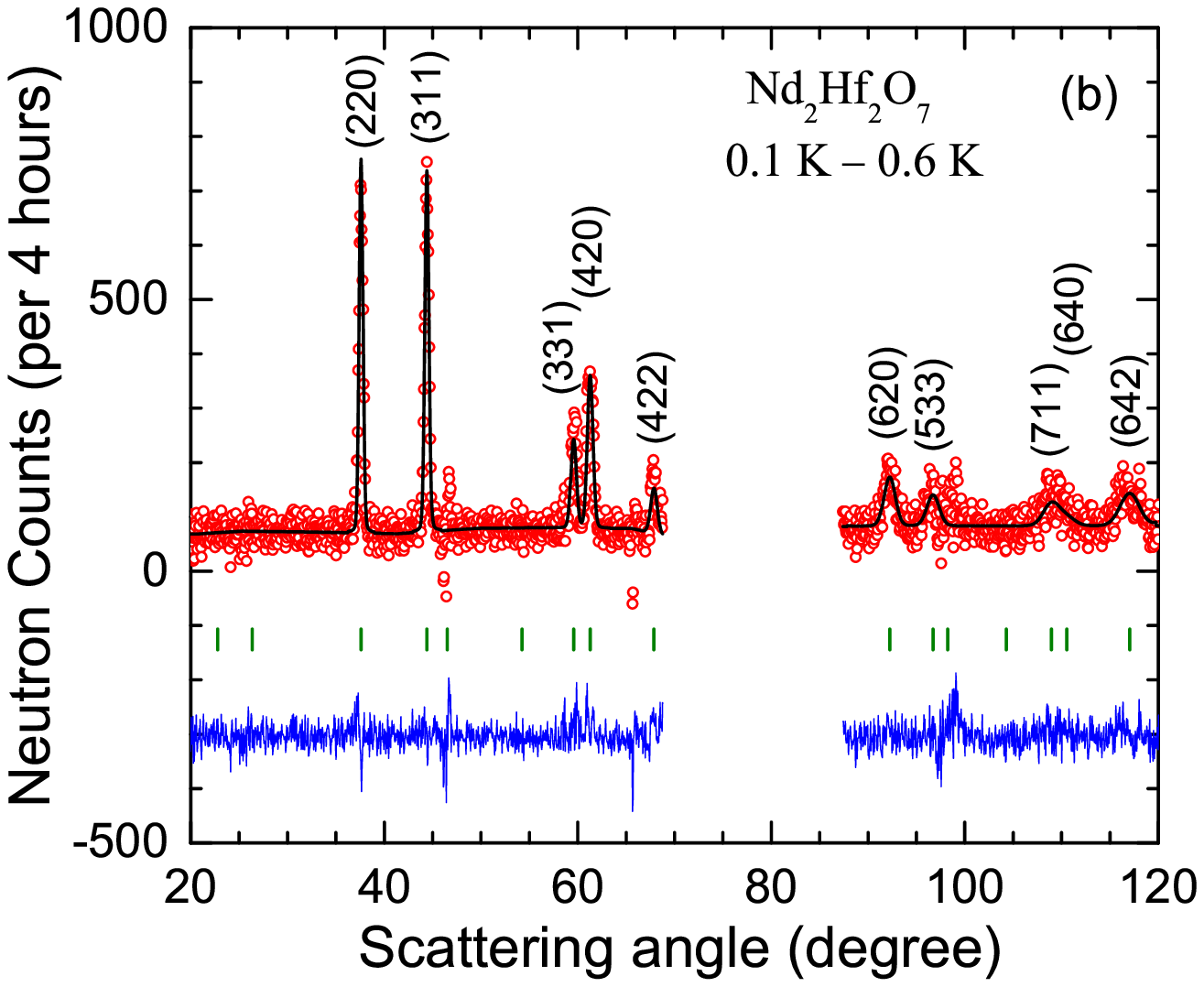}
\caption {(Color online) (a) Neutron diffraction (ND) pattern of Nd$_2$Hf$_2$O$_7$ recorded at 0.6~K\@. The solid line through the experimental points is the calculated pattern by considering the ${\rm Eu_2Zr_2O_7}$-type face centered cubic (space group $Fd\bar{3}m$) pyrochlore nuclear structure. The short vertical bars mark the Bragg peak positions of primary phase [upper row (olive)] and sample holder Cu [lower row (orange)]. The lowermost curve represents the difference between the experimental and calculated patterns. The two most intense peaks marked with stars ($\star$) belong to the sample holder. Inset: Expanded view of ND pattern between 35$^\circ$--45$^\circ$ and comparison of ND patterns at 0.6~K and 0.1~K to highlight the presence of magnetic scattering (marked with \#). (b) Magnetic diffraction pattern at 0.1~K (after subtracting 0.6~K nuclear pattern) together with the calculated magnetic refinement pattern. The region where the sample holder contribution dominates is excluded. The difference between the experimental and calculated intensities is shown by the blue curve at the bottom. The Miller indices ($hk\ell$) of the strongest magnetic Bragg peaks are indicated.}
\label{fig:ND}
\end{figure}

\begin{figure}
\includegraphics[width=2.8in]{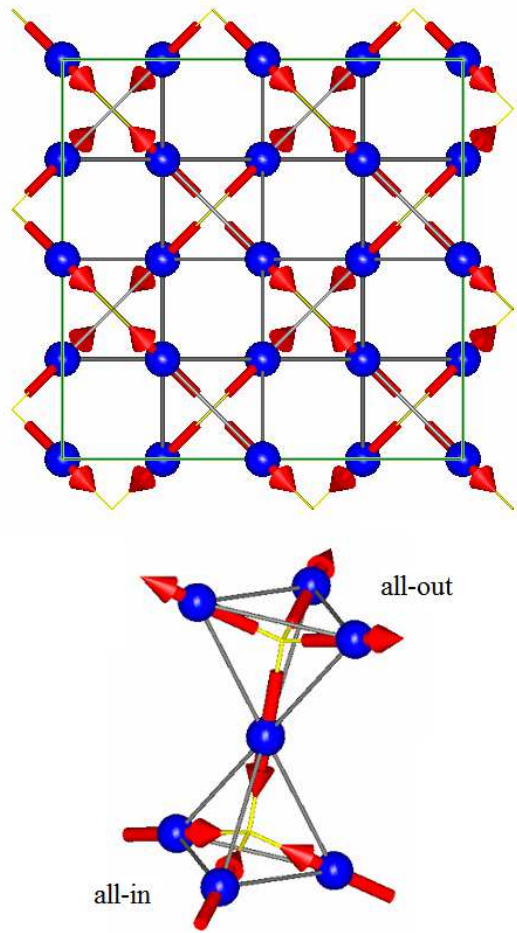}
\caption {(Color online) A two dimensional view of all-in/all-out magnetic structure of Nd$_2$Hf$_2$O$_7$ along with the three dimensional view of corner-shared `all-in' and `all-out' tetrahedra. The arrows denote the odered Nd$^{3+}$ moment directions, pointing towards or away from the center of tetrahedra. The two dimensional representation of crystal structure can be seen in Fig.~\ref{fig:struct}(b) and the three dimensional view of corner-shared tetrahedra in Fig.~\ref{fig:struct}(a).}
\label{fig:MagStruct}
\end{figure}

The neutron diffraction data collected at 0.6 K  are shown in Fig.~\ref{fig:ND}(a) together with the calculated pattern for the nuclear structure of  Nd$_2$Hf$_2$O$_7$. The crystallographic parameters are listed in Table~\ref{tab:XRD} and are consistent with the parameters from room temperature XRD. This also confirms that there is no structural transition down to 0.6~K\@. At 0.1~K, weak and noticeable additional intensities in the diffraction pattern confim the long range magnetic ordering. A comparison of the ND data collected at 0.6~K ($> T_{\rm N}$) and 0.1~K ($< T_{\rm N}$) is shown in the inset of Fig.~\ref{fig:ND}(a) for $2\theta$ range 35$^\circ$--45$^\circ$. An enhancement of the two nuclear peaks (2\,2\,0) and (3\,1\,1) can be seen (marked with symbol \#). This additional intensity corresponds to the most prominent magnetic peaks in the ordered state. From the difference between the ND patterns recorded at 0.1~K and 0.6~K [Fig.~\ref{fig:ND}(b)] a multitude of magnetic Bragg peaks can be clearly visible. The appearance of an additional magnetic Bragg peak (4\,2\,0) at $59.6^\circ$  confirms the antiferromagnetic ordering. All magnetic Bragg peaks are well indexed by the propagation wavevector {\bf k} = (0, 0, 0). The propagation wavevector {\bf k} = (0, 0, 0) was also found to index the magnetic Bragg peaks in Nd$_2$Zr$_2$O$_7$  \cite{Lhotel2015,Xu2015}.

In order to determine the magnetic structure compatible with the space group symmetry we carried out representational analysis using the program BASIREPS from the FullProf package \cite{Rodriguez1993}. The symmetry analysis for the propagation vector {\bf k} = (0, 0, 0) and space group $Fd\bar{3}m$ yielded four nonzero irreducible representations (IRs) for the magnetic Nd(16d) site: 1 one-dimensional ($\Gamma_3^1$), 1 two-dimensional ($\Gamma_6^2$) and two three-dimensional ($\Gamma_8^3$, $\Gamma_{10}^3$)  for the little group. The magnetic representation $\Gamma_{\rm mag\,Nd}$ is thus composed of four IRs as
\begin{equation}
\Gamma_{\rm mag\,Nd} = 1\,\Gamma_3^1 + 1\,\Gamma_6^2 + 1\,\Gamma_8^3 + 2\, \Gamma_{10}^3.
\label{eq:IRs}
\end{equation}
The IRs $\Gamma_3^1$, $\Gamma_6^2$ and $\Gamma_8^3$ enters only once in magnetic decomposition whereas $\Gamma_{10}^3$ is repeated twice. The basis vectors (BVs) of these IRs are listed in Table~\ref{tab:ND}. While the BVs vector of $\Gamma_6^2$ consist of both real and imaginary components, the BVs for $\Gamma_3^1$, $\Gamma_8^3$ and $\Gamma_{10}^3$ have only real components. As listed in Table~\ref{tab:ND}, $\Gamma_3^1$ has one BV, $\Gamma_6^2$ has two, $\Gamma_8^3$ has three and $\Gamma_{10}^3$ has six BVs.

\begin{table*}
\caption{\label{tab:ND} Nonzero irreducible representations (IRs) and associated basis vectors $\psi_\nu$ for Nd(16d) site in space group $Fd\bar{3}m$ with propagation vector {\bf k} = (0, 0, 0) for  Nd$_2$Hf$_2$O$_7$ obtained from the representational analysis using the program BASIREPS.  The atoms of the nonprimitive basis are defined according to Nd1: (0.50, 0.50, 0.50); Nd2: (0.25, $-$0.25, 1.00); Nd3: ($-$0.25, 1.00, 0.25); Nd4: (1.00, 0.25, $-$0.25).}
\begin{ruledtabular}
\begin{tabular}{lcccccc}
IRs & $\psi_\nu$ & component & Nd1 & Nd2 & Nd3 & Nd4 \\
 \hline
$\Gamma_3^1$ & $\psi_1$ & Real &(1\,1\,1) & ($-$1\,$-$1\,1) & ($-$1\,1\,$-$1) & (1\,$-$1\,$-$1) \\

$\Gamma_6^2$ & $\psi_1$ & Real      & (1\,$-$0.5\,$-$0.5)& ($-$1\,0.5\,$-$0.5) &  ($-$1\,$-$0.5\,0.5) & (1\,0.5\,0.5) \\
             &          & Imaginary & (0\,$-$0.87\,0.87) & (0\,0.87\,0.87) & (0\,$-$0.87\,$-$0.87) & (0\,0.87\,$-$0.87) \\
             & $\psi_2$ & Real      &($-$0.5\,1\,$-$0.5) & (0.5\,$-$1\,$-$0.5) & (0.5\,1\,0.5) & ($-$0.5\,$-$1\,0.5) \\
             &          & Imaginary &(0.87\,0\,$-$0.87) & ($-$0.87\,0\,$-$0.87) & ($-$0.87\,0\,0.87) & (0.87\,0\,0.87) \\
$\Gamma_8^3$ & $\psi_1$ & Real      & (1\,$-$1\,0)& ($-$1\,1\,0)& (1\,1\,0) & ($-$1\,$-$1\,0)\\
             & $\psi_2$ & Real      & (0\,1\,$-$1)& (0\,1\,1) & (0\,$-$1\,$-$1) & (0\,$-$1\,1)\\
             & $\psi_3$ & Real      & ($-$1\,0\,1) & ($-$1\,0\,$-$1) & (1\,0\,-1)& (1\,0\,1)  \\	
$\Gamma_{10}^3$ & $\psi_1$ & Real     & (1\,1\,0) & ($-$1\,$-$1\,0) & (1\,$-$1\,0)& ($-$1\,1\,0)\\
             & $\psi_2$ & Real      & (0\,0\,1) & (0\,0\,1) & (0\,0\,1) & (0\,0\,1)\\
             & $\psi_3$ & Real      & (0\,1\,1) & (0\,1\,$-$1) & (0\,$-$1\,1) & (0\,$-$1\,$-$1)\\
             & $\psi_4$ & Real      & (1\,0\,0)&  (1\,0\,0)& (1\,0\,0) & (1\,0\,0)\\
             & $\psi_5$ & Real      & (1\,0\,1)& (1\,0\,$-$1)& ($-$1\,0\,$-$1)& ($-$1\,0\,1)\\
             & $\psi_6$ & Real      & (0\,1\,0) & (0\,1\,0) & (0\,1\,0) & (0\,1\,0) \\
\end{tabular}
\end{ruledtabular}
\end{table*}

Out of the above four IRs the best refinement of the magnetic diffraction pattern is obtained for $\Gamma_3^1$ (with a magnetic $R$ factor of 10.1\%) which corresponds to the all-in/all-out spin configuration. For the refinement, the scale factor was fixed to the value obtained from the nuclear structure refinement at 0.6~K\@. Only the coefficient of one basis vector of $\Gamma_3^1$ was the refinable parameter. The refinement of the magnetic diffraction pattern at 0.1~K with the all-in/all-out type model as shown in Fig.~\ref{fig:ND}(b) gives an ordered moment of $m = 0.62(1) \,\mu_{\rm B}$/Nd. The all-in/all-out magnetic structure of Nd$_2$Hf$_2$O$_7$ is illustrated in Fig.~\ref{fig:MagStruct}. The magnetic structure is comprised of alternating `all-in' and `all-out' units of corner-shared tetrahedra, where each tetrahedral unit consists of four Nd$^{3+}$ magnetic moments at the vertices of the tetrahedra all pointing either towards the center (all-in) or away from the center (all-out) of tetrahedra as illustrated in the lower panel of Fig.~\ref{fig:MagStruct}.

\begin{figure} [b]
\includegraphics[width=3in]{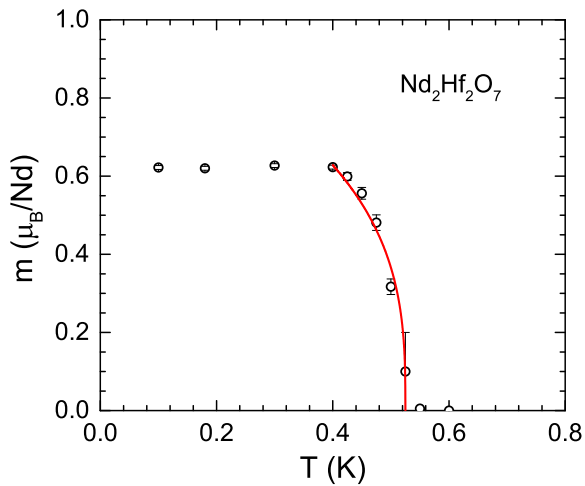}
\caption{\label{fig:NDmoment} (Color online) Temperature $T$ dependence of the ordered moment $m$ (per Nd ion) obtained from the refinement of neutron powder diffraction data at various temperatures. The solid curve represents the fit according to $m = m_0 (1 - T/T_{\rm N})^{\beta}$ for $T_{\rm N} = 0.53(1)$~K and $\beta=0.34(5)$ in \mbox{0.4~K~$\leq T \leq T_{\rm N}$}.}
\end{figure}

The $T$ dependence of the ordered moment $m$ obtained from the refinement of ND patterns at different temperatures is shown in Fig.~\ref{fig:NDmoment}. The $m(T)$ data above $T= 0.4$~K are reasonably described by $m = m_0 (1 - T/T_{\rm N})^{\beta}$. The fit for 0.4~K~$\leq T \leq T_{\rm N}$ is shown by the solid curve in Fig.~\ref{fig:NDmoment} which gives $T_{\rm N} = 0.53(1)$~K and $\beta=0.34(5)$. The critical exponent $\beta$ is close to $ \beta \approx 0.33$ for a three-dimensional Ising system \cite{Blundell2001}. We also notice that $m(T)$ at $T \leq 0.4$ K is almost independent of $T$ with a value of  $\sim 0.62\,\mu_{\rm B}$/Nd which is quite unusual. The origin of this unusual flattening of $m(T)$ is not clear and requires further investigation. Because of the anisotropic nature one would expect a gapped magnon spectrum in this compound, however a small energy gap will not be sufficient to explain the observed $T$ dependence of $m(T)$ at $T \leq 0.4$~K\@.

For an effective $S = 1/2$ with effective $g_{zz} = 5.01(3)$ one would expect an ordered moment of $g_{zz} S = 2.50 \, \mu_{\rm B}$/Nd. However, the obtained ordered moment $0.62(1)\,\mu_{\rm B}$/Nd at 0.1~K is much smaller than this value. The strongly reduced value of ordered moment reflects the presence of strong quantum fluctuations persisting deep into the ordered state down to 0.1~K\@. In a pyrochlore material with local $\langle 111 \rangle$ Ising anisotropy the magnetic ground state strongly depends on the relative strength of the dipolar and magnetic exchange interactions \cite{Bramwell2001,Hertog2000}. Spin ice behavior is observed when the ferromagnetic dipolar interaction dominates over the antiferromagnetic exchange. On the other hand a dominating antiferromagnetic exchange stabilizes all-in/all-out long range magnetic order. We estimate the nearest neighbor dipole-dipole interaction $D_{\rm nn}$ using our effective moment $\mu_{\rm eff} = 2.45\, \mu_{\rm B}$/Nd and unit cell lattice parameter $a = 10.6389(1)$~\AA\ as \cite{Bramwell2001,Hertog2000}
\begin{equation}
D_{\rm nn} = \frac{5}{3}\left(\frac{\mu_0}{4\pi}\right) \frac{\mu_{\rm eff}^2}{r_{\rm nn}^3} \approx 0.12~{\rm K}
\label{eq:Dnn}
\end{equation}
where $\mu_0$ is the magnetic permeability of vacuum and $r_{\rm nn }= (a/4)\surd 2$ is the nearest neighbor distance. Following Siddharthan {\it et al}.\ \cite{Siddharthan1999} a rough estimate of nearest neighbor exchange interaction between the $\langle 111 \rangle$ Ising moments $J_{\rm nn}$ can be made by fitting the $\chi(T)$ data with a high-temperature series expansion up to order $(1/T^2)$, i.e. by  $\chi(T) = (C_1/T)[1+C_2/T]$ where $C_2$ can be decomposed as a sum of exchange and dipolar interactions, $C_2 = (6S^2/4)[2.18 D_{\rm nn} + 2.67 J_{\rm nn}]$. The fit of $\chi(T)$ data with this expression in 10~K~$\leq T\leq 30$~K gives $C_2 = -0.67(5)$~K. Thus using the above estimated $D_{\rm nn} \approx 0.12$~K, from $C_2$ we obtain $J_{\rm nn} \approx -0.77$~K\@.  Though this estimate of $J_{\rm nn}$ is not very accurate, it clearly shows that the antiferromagnetic $J_{\rm nn}$ dominates over the dipolar $D_{\rm nn}$, eventually leading to a long range ordered ground state in Nd$_2$Hf$_2$O$_7$ with an all-in/all-out magnetic structure. A better estimation of $J_{\rm nn}$ is desired to check if the ratio $J_{\rm nn}/D_{\rm nn}$ complies with the phase diagram of Ising pyrochlore magnets which predicts a long range antiferromagnetic ordering for $J_{\rm nn}/D_{\rm nn} < -0.91$ \cite{Hertog2000}. We would also like to point out that because of the octupolar tensor component for Nd$^{3+}$ \cite{Huang2014}, the estimate of $D_{\rm nn}$  using Eq.~(\ref{eq:Dnn}), which only accounts for the dipolar component, may also not be very precise.

\section{\label{Conclusion} Summary and Conclusions}

The physical properties of a pyrochlore hafnate Nd$_2$Hf$_2$O$_7$ have been investigated by $\chi_{\rm ac}(T)$, $\chi(T)$, $M(H)$ and $C_{\rm p}(T)$ measurements. Evidence of an antiferromagnetic transition below $T_{\rm N} =0.55$~K is seen from the $T$ dependence of $\chi_{\rm ac}$ measured down to 0.2~K\@. The analysis of $M(H)$ data indicate a local $\langle 111 \rangle$ Ising anisotrpic behavior with an effective longitudinal $g$-factor of 5.01(3) for the pseudo spin-half Kramers doublet ground state of Nd$^{3+}$. The low-$T$ $\chi(T)$ reveals an effective moment of $2.45\, \mu_{\rm B}$/Nd for the Ising ground state and a positive $\theta_{\rm p}$ reflecting ferromagnetic coupling between the Nd spins, though the compound orders antiferromagnetically. The $C_{\rm p}(T)$ data show the presence of short range correlations well above the antiferromagnetic ordering. The crystal field analysis of $C_{\rm mag}(T)$ suggests the splitting energy between the ground state doublet and first excited state doublet to be about 230~K\@.

Magnetic structure determination by neutron powder diffraction confirmed the long range antiferromagnetic ordering with a magnetic propagation wavevector {\bf k} = (0, 0, 0). The  Nd$^{3+}$ moments are found to adopt an all-in/all-out structure with the four magnetic moments at the vertices of the tetrahedra pointing alternatively either all-into or all-out-of the centers of the neighboring tetrahedra. The ordered state magnetic moment of Nd$^{3+}$ $m = 0.62(1)\,\mu_{\rm B}$/Nd at 0.1~K is highly reduced compared to the expected Ising moment value of $2.50\,\mu_{\rm B}$/Nd with an effective spin $S = 1/2$ and $g_{zz} = 5.01(3)$ Kramers doublet ground state, reflecting the presence of strong quantum fluctuations. The unusual reduction of ordered moment and presence of strong quantum fluctations could be due to the dipolar-octupolar nature of Kramers doublet ground state  of Nd$^{3+}$ \cite{Huang2014}, which however remains to be confirmed by further theoretical and experimental  works.

\acknowledgements
We acknowledge Helmholtz Gemeinschaft for funding via the Helmholtz Virtual Institute (Project No. VH-VI-521).

\end{document}